\pdfoutput=1
\documentclass[journal]{IEEEtran}

\usepackage{amsmath}
\usepackage{pgfplots}
\usetikzlibrary{arrows, calc, patterns, shapes}
  \pgfplotsset{compat=1.15}
\usepackage{fouriernc}
\usepackage{graphicx}
\usepackage{glossaries}
\usepackage{graphicx}
\makeglossaries
\ifCLASSINFOpdf
\else
\fi
\begin{document}
\title{Quaternion-Aligned Electroanatomical Models of the Murine Atria from Standard-Output Functional and Structural Imaging}
\author{Gavin Tumlinson, Enaam Chleilat, Josef Madl, Peter Kohl, Viviane Timmermann}

%
%

\markboth{ }%
{Shell \MakeLowercase{\textit{et al.}}: Bare Demo of IEEEtran.cls for IEEE Journals}
%



\maketitle
\newglossaryentry{AF}{name = {AF}, description = {atrial fibrillation}}
\newglossaryentry{2D}{name = {2D}, description = {two dimensional}}
\newglossaryentry{3D}{name = {3D}, description = {three dimensional}}

\IEEEpeerreviewmaketitle

\begin{abstract}
    
Our evolving understanding of the heterocellular cardiac environment demands innovative tools for its study. While murine models are lauded for their versatility and accessibility, they are constrained by scale; tools designed for larger subjects are unsuitable in mice. Advanced techniques, such as optical trans-membrane potential mapping and three-dimensional microscopy may fill this void. To enhance the utility of these methods, we developed a novel cross-correlation technique grounded in quaternion mathematics. The successful integration of multimodal observations represents a significant advancement in our ability to study cardiac electrophysiology in murine models, providing insight into mechanisms previously unexplored due to technical constraints.

\end{abstract}
\section{Introduction}
\IEEEPARstart{C}{ardiac} electrophysiology has long considered myocytes as the heart’s sole electrically active cell type. Yet, emerging evidence of non-myocytes engaging in electrotonic coupling with myocytes, most notably at injury borders, challenges this notion \cite{quinn2016electrotonic}.

Such insights are crucial in clinical settings, particularly for interventions like \gls{AF} ablation, where therapeutic strategies often hinge on disrupting aberrant electrical pathways. Despite the efficacy of such procedures, long-term success is variable  \cite{jais2003radiofrequency}, highlighting the potential role of alternative substrates for conduction through scar tissue. Therefore, steering the cardiac electrical environment requires a holistic understanding of both participating cellular actors and the underlying substrate — especially at and around injury sites.

While mouse models are valuable for exploring scar formation, they fall short in the facile generation and precise assessment of lesions. Techniques like optical mapping of the curved \gls{2D} cardiac surface and high-resolution \gls{3D} microscopy offer profound insights but do not allow simultaneous functional and structural integration. Interlinking these outputs is no small feat, especially when aiming for a harmonious and comprehensive model. Our method endeavours to bridge this gap, ultimately aiming to lay a robust foundation upon which precise electrophysiological simulations can be built.
\section{System Model}
To replicate \gls{AF} ablation lesions, mice underwent surgical or cryoablation of the left atrial auricule. At 28 or 56 days post-procedure, they were euthanized for heart extraction and Langendorff perfusion with a voltage-sensitive dye; subsequently, the atria were dissected and superfused for single-viewpoint optical mapping of membrane potentials. Thereafter, the samples were fixed in 4\% paraformaldehyde, passively cleared with the X-CLARITY technique \cite{chung2013structural}, and stained for collagen, alpha-actinin, and cell nuclei. High-resolution
(0.1µm x 0.1 µm x 50µm per voxel) \gls{3D} imaging of the atria was performed using a Leica TCS SP8 DIVE multiphoton microscope.

Both methodologies face inherent imaging challenges. Confocal microscopy, optimized for in-plane scanning, can compromise depth resolution due to sample stability concerns in lengthy runs. Optical mapping balances temporal and spatial resolution, but is limited by fluorescence intensity — especially in the much thinner atrial wall — photobleaching, and its quasi-\gls{2D} information content.

Structure and function are often not immediately congruent; our microscopy images do not project seamlessly onto the corresponding optical maps. While several factors may contribute, marked variation between functional recordings of the same sample supports the notion that unintended repositioning — collapsed in \gls{2D} — is a source of observed discrepancy.

Most structural features cannot be conclusively linked to analogous nuances in optical mapping. Even time-independent deviations in apparent brightness may be attributable to heterogeneous dye loading and distribution of excitable tissue. However, the sample outline is reliable in the optical maps and can be matched to a comparable contour from an axially collapsed microscopy dataset.

\section{Cross-correlation}

To correlate structural 3D confocal microscopy images with 2D optical mapping images, we developed a novel algorithm to identify the projection axis and rotation of the 3D model where it maximally aligns with the 2D observation. A smoothed surface hull  $z = \mathbf{O}(x,y)$ was approximated from discrete pixels in the structural dataset with bicubic spline interpolation. The 3D model is projected onto a 2D plane at varying rotation about the fixed centroid, yielding a 1D function of its contour, and aligned with the likewise interpolated optical map. Alignment was evaluated by multiplying and integrating the contour functions; when similar the area under the curve is maximal and the datasets most congruent. The optimum axis and rotation are found using a modified Davidon-Fletcher-Powell approach to maximizing the previously determined alignment.








Discrete microscopy images of the murine atrium can be approximated as a complex 3D structure, encapsulated by a surface hull $ z = \mathbf{O}(x,y)$, using bicubic spline interpolation (for detailed reading refer to \cite{numrecfort77}). The resultant 2D projection  $\mathbf{P}$ is conceptualized as the integral of $\mathbf{O}$ with respect to the projection axis \( \hat{n} \). Here, quaternion conjugation is particularly useful for describing 3D rotation. A quaternion \(q)\) is a four-dimensional versor, with a real and vector part. An optimal quaternion \( q^{*}\) with vector part $\vec{v}^{*}=(b^{*},c^{*},d^{*})\in \mathbb{R}^3$, is sought, that upon conjugation with $z = \mathbf{O}(x,y)$ and projection along \( \hat{n} = \frac{\vec{v^*}}{||\vec{v^*}||} \), maximizes congruence with the optical mapping contour $c_{OM}(u,p) \in \mathbf{P}^*$. The projection operation translates to integrating over lines parallel to \( \hat{n} \):
\begin{equation}
\mathbf{P}^*(u, p) = \int \mathbf{O}(x(\Bar{u}, \Bar{p}, \Bar{n}), y(\Bar{u}, \Bar{p}, \Bar{n})), z(\Bar{u}, \Bar{p}, \Bar{n}))) \, d\Bar{n}
\end{equation}
where  \( (\Bar{u}, \Bar{p},) \) are scalars of \( \hat{u} \) and \( \hat{p} \) respectively, and \( \Bar{n} \) serves as a scalar to parameterize the position along \( \hat{n} \). Considering \( \hat{u} \) and \( \hat{p} \) as normalized orthogonal unit vectors \( \in \mathbf{P}^*\) , $ (\Bar{u}, \Bar{p}, \Bar{n})$ can be translated to the coordinates $(x,y,z)$:

\begin{equation}
    \begin{split}
    x(\Bar{u}, \Bar{p}, \Bar{n}) = \Bar{u} \hat{u}_x + \Bar{v}  \hat{p}_x + \Bar{n}  \hat{n}_x\\
    y(\Bar{u}, \Bar{p}, \Bar{n}) = \Bar{u} \hat{u}_y + \Bar{v}  \hat{p}_y + \Bar{n}  \hat{n}_y\\
    z(\Bar{u}, \Bar{p}, \Bar{n}) = \Bar{u} \hat{p}_z + \Bar{v} \hat{p}_z + \Bar{n}  \hat{n}_z
    \end{split}
\end{equation}

$ z = \mathbf{O}(x,y)$ condenses into a piecewise differentiable closed loop $c_z(x,y)$ due to possible concavities that do not contribute to the 2D projection.

Its alignment with the optical mapping sample outline $c_{OM}(\Bar{u},\Bar{p}) \in \mathbf{P}^*$, is evaluated by summing the integral of overlap of each differentiable segment of $c^*(\Bar{u},\Bar{p})$ and $c_{OM}(\Bar{u},\Bar{p}$):

\begin{equation}
    \sum_i \oiint_{S_i} c^*(\Bar{u},\Bar{p}) \cdot c_{OM}(\Bar{u},\Bar{p}) \, d\Bar{u}, d\Bar{p}
\end{equation}

where $S_i$ denotes the domain of the $i$-th segment. A segment is a shared differentiable domain of both $c^*(\Bar{u},\Bar{p})$ and $c_{OM}(\Bar{u},\Bar{p}$), which ends when either spline reaches its border or a singularity resulting from projection is encountered. 

The construction of \( z = \mathbf{O}(x, y) \) greatly impacts the efficiency of calculating the local ideal \( q* \). Bicubic spline interpolation not only facilitates a smoothed representation of the structure but also enhances the computational efficiency in solving the maximization problem for finding an optimal quaternion \( q^* \):
\begin{equation}
    q^{*} = \underset{\mathbf{q}}{\text{argmax}} \, (\sum_i \oiint_{S_i} c^{*}(u,v; q) \cdot c_{OM}(u,v) \,du dv)
\end{equation}  
Piecewise continuity is pivotal here. It enables the computation of changes in the projection caused by component variations in \(q\) without reevaluating at each iterative adjustment. Sections can be omitted from calculation if they do not contribute to the projection, particularly when they are concave in the direction of integration or border a further outlying convex segment that preempts their inclusion in it. Moreover, since the partial derivatives with respect to each component of \( q \) are calculable, and interpolation across quaternion rotations is feasible, local derivatives need to be calculated only once per spline, and can then be adapted by conjugation with the component adjustment of \( q \). Setting the gradient \( \nabla_q \) with respect to its scalar components to zero:
\begin{equation}
    \nabla_q (\sum_i \oiint_{S_i} c^{*}(\Bar{u},\Bar{p}; q) \cdot c_{OM}(\Bar{u},\Bar{p}) \, d\Bar{u} d\Bar{p}) = 0
\end{equation}
 
yields a system of equations that, when solved, provides the local critical points of the alignment function. Though not always necessary, iterating through a wide combination of intersected border splines further solidifies the assurance that the found \( q^* \) is indeed a good approximation of the optimum.

Once aligned along \( q^* \), superimposing the datasets is straightforward. From this perspective, the functional data can be cast without translation onto the smoothed structural model and interpreted in context.

\begin{figure}
    \centering
    \includegraphics[width=0.5\textwidth]{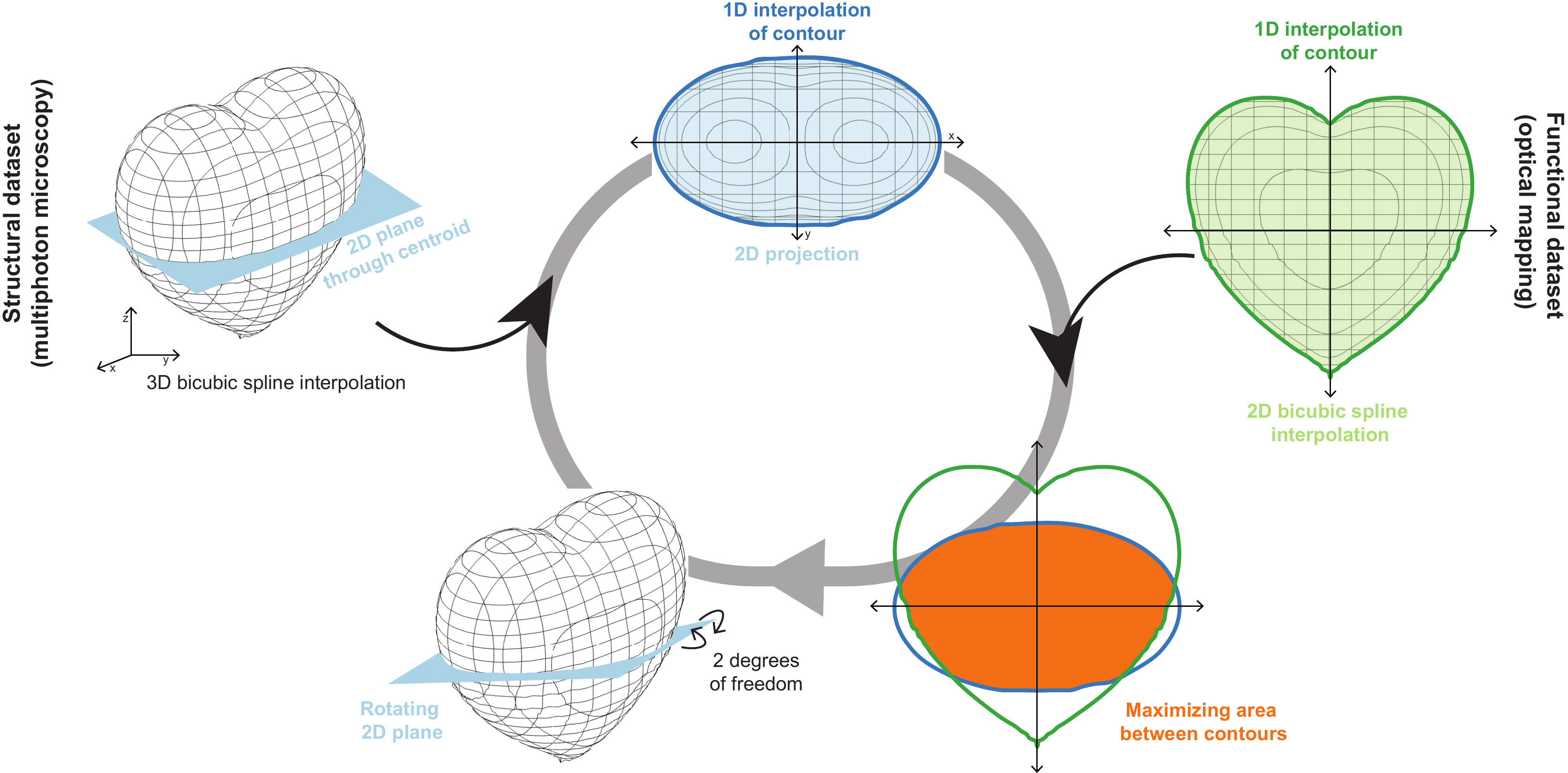}
    \caption{\textbf{Schematic of the method for determining the optimal 2D plane that intersects the approximated 3D multiphoton data, and achieves maximum alignment with the 2D contour of the optical mapping data.} [top, left] We approximate the surface of the structural 3D multiphoton data using bicubic splines. Then, we generate an arbitrary 2D plane [light blue] that intersects with the centroid of the 3D object. [top, middle] The object is projected onto the 2D plane, and the contour of the projection is approximated by a 1D function in dark blue. [top, right] The contour of the 2D functional optical mapping data [light green] is approximated by a 1D function in dark green. [bottom, right] The alignment between the contour functions of the multiphoton data [dark blue] and the optical mapping data [dark green] is compared. [bottom, left] The 2D plane, anchored at the centroid, undergoes rotation via a modified Davidon-Fletcher-Powell approach to optimise the alignment between the contour of the multiphoton data and optical mapping data.}
    \label{fig:enter-label}
\end{figure}
\section{Discussion}

We acknowledge that deformation during sample preparation and clearing remains unaccounted for in our current implementation, which could influence data interpretation. Additionally, applying a rolling Fourier transform across the model's surface — \( F(\omega) = \int_{-\infty}^{\infty} f(t) e^{-i\omega t + \delta} dt \), with \(\delta\) varying based on signal vector and depth — may provide a refined method for extracting wave-like excitations from the integrated structure-function data.
\section{Conclusion}
In addressing the spatial mismatch between structural and functional imaging information, our approach surpasses traditional limitations. Future research could enhance this method; incorporating machine learning and dynamic artefact exclusion may further improve accuracy. Advanced analytical techniques hold promise for new insights into finer aspects of cardiac electrical conduction and anomalies.
\section*{Acknowledgment}
The authors extend their sincere gratitude to Dr. Callum Michael Zgierski-Johnston, Thomas Kok, Jonas Heer, and Alyssa Burgueño, whose assistance has been instrumental in the developing of this method.

\ifCLASSOPTIONcaptionsoff
  \newpage
\fi

\end{document}